\documentclass[twocolumn,showpacs,preprintnumbers,amsmath,amssymb]{revtex4}
\usepackage{epsfig}
\usepackage{dcolumn}
\usepackage{bm}
\def\bk{\hspace{-5pt}/}

\begin{document}

\preprint{CERN-PH-TH/2009-146}

\title{Origin and Phenomenology of Weak-Doublet Spin-1 Bosons}

\author{M. V. Chizhov$^{1,2}$, Gia Dvali$^{3,4,5}$}
 \affiliation{$^1$ \mbox{CSRT, Faculty of Physics, Sofia University, 1164 Sofia, Bulgaria}\\
$^2$ \mbox{DLNP, Joint Institute for Nuclear Research, 141980, Dubna, Russia}\\
$^3$ \mbox{CERN, Theory Unit, Physics Department, CH-1211 Geneva 23, Switzerland}\\
$^4$ \mbox{CCPP, Department of Physics, New York University, 4
Washington Place, New York, NY 10003, USA}\\
$^5$ \mbox{Max-Planck-Institute for Physics, F\"ohringer  Ring 6,
D-80805, M\"unchen,  Germany}}


\begin{abstract}
We study phenomenological consequences of the Standard Model
extension by the new spin-1 fields with the internal quantum numbers
of the electroweak Higgs doublets. We show, that there are at least
three different classes of theories, all motivated by the hierarchy
problem, which predict appearance of such vector weak-doublets not
far from the weak scale. The common feature for all the models is
the existence of an $SU(3)_W$ gauge extension of the weak $SU(2)_W$
group, which is broken down to the latter at some energy scale
around TeV. The Higgs doublet then emerges as either a
pseudo-Nambu-Goldstone boson of a global remnant of $SU(3)_W$, or as
a symmetry partner of the true eaten-up Goldstone boson.  In the
third class, the Higgs is a scalar component of a high-dimensional
$SU(3)_W$ gauge field. The common phenomenological feature of these
theories is the existence of the electroweak doublet vectors
$(Z^*,W^*)$, which in contrast to well-known $Z'$ and $W'$ bosons
posses only anomalous (magnetic moment type) couplings with ordinary
light fermions. This fact leads to some unique signatures for their
detection at the hadron colliders.
\end{abstract}


\maketitle

\section{Introduction}

The main theoretical motivation for beyond the standard model
physics around TeV energies is provided by the Hierarchy Problem,
inexplicable quantum stability of the weak interaction scale with
respect to the ultraviolet cutoff. This problem suggests the
existence of some new regulating physics not far above the weak
scale.  Needless to say, understanding experimental consequences of
the latter is of  fundamental importance.

Recently~\cite{misha}, it was pointed out that possible existence of
massive vector fields ($V_{\mu} \, \equiv (Z^{*}_{\mu},
W_{\mu}^{*-})$) with the internal quantum numbers identical to the
Standard Model Higgs (or the lepton) doublet, can result in some
interesting phenomenological consequences.  Due to their quantum
numbers, to the leading order such vectors can only have magnetic
type interactions with the Standard Model fermions,
\begin{eqnarray}
\nonumber {1 \over M}  D_{[\mu} V_{\nu]}^{\rm c}  \, \left (
g^d_{LR} \, \bar{Q}_L \sigma^{\mu\nu} d_R \,
+ \, g^e_{LR} \, \bar{L}\sigma^{\mu\nu} e_R  \right )  \\
+{g^u_{LR} \over M}  D_{[\mu} V_{\nu]} \,   \bar{Q}_L
\sigma^{\mu\nu} u_R \, +{\rm h.c.}, \label{maincoupling}
\end{eqnarray}
where $V_{\mu}^{\rm c} \, \equiv (-W_{\mu}^{*+}, \bar{Z}^{*}_{\mu})$
is the charge-conjugated doublet; $Q_L\equiv (u_L, d_L)$ and $L
\equiv (\nu_L, e_L) $ are the left-handed quark and lepton doublets
respectively. $D_{\mu}$ are the usual $SU(2)_W\times U(1)_Y$
covariant derivatives, and the obvious group and family  indexes are
suppressed. $M$ is the scale of the new physics and $g^{u,d,e}_{LR}$
are dimensionless constants.

Up until now, no theoretical motivation for the existence of such
states was given. It is the purpose of this paper to provide such a
motivation from the Hierarchy Problem point of view. We shall show
that such states are predicted by three different classes of
theories that represent different approaches for explaining the
relative lightness of the Higgs doublets.

The crucial common feature of all three approaches  is, that they
are based on the existence of $U(3)_W \, \equiv \, SU(3)_W\times
U(1)_W $ gauge extension of the $SU(2)_W\times U(1)_Y$ electroweak
group , which is spontaneously broken down to the latter at scale
$M$.  The weak doublet vectors $V_{\mu}$ are then identified with
the $SU(2)_W$ doublet components of the $8$-dimensional gauge
multiplet of the $SU(3)_W$ group, which under $SU(2)_W$ subgroup
decomposes as
\begin{equation}
\label{3to2}
 {\bf 8 \, = \, 3 \, + \, 2\, +\, \bar{2}\, +\, 1},
\end{equation}
where numbers refer to the dimensionality of the corresponding
$SU(2)_W$ representations. The vector fields $V_{\mu}$
($V_{\mu}^\dagger$) obviously belong to fragments ${\bf 2}$ (${\bf
\bar{2}}$), and become massive during the spontaneous symmetry
breaking $ U(3)_W \, \rightarrow \, SU(2)_W \times U(1)_Y$. The
lightness of the Higgs doublets is guaranteed, because they are
related to $V_{\mu}$ vectors by symmetry. This relation in three
different approaches is established as follows.

\subsection{Pseudo-Goldstone Higgs}

In the first approach the lightness of the Standard Model Higgs
doublet is achieved because it is a  pseudo-Goldstone boson of a
spontaneously broken $G_{global} \, \equiv \, G \times G$ symmetry
of the scalar potential, whereas only the diagonal $G_{local} \,
\equiv \, G$ part of it is gauged.  This idea was originally
proposed in the context of $G = SU(6)$ grand unification~\cite{u6},
but  our current focus will be  the $G = U(3)_W$ realization of  it
\cite{u3}. In the latter  realization  $SU(2)_W\times U(1)_Y$ is
embedded into $G_{local} = U(3)_W$ group as a maximal subgroup.   As
explained above, the $8$-dimensional gauge multiplet of $SU(3)_W$,
on top of the $3$ electroweak gauge bosons and an extra singlet,
contains a complex doublet $V_{\mu}$ with the quantum numbers of the
Standard Model Higgs.

The spontaneous breaking of the local symmetry $U(3)_{local}
\rightarrow SU(2)_W\times U(1)_Y$ is triggered by the two
independent Higgs triplets, ${\bf 3}_H$ and ${\bf 3}_H'$, which
under the $SU(2)_W$ subgroup decompose as ${\bf 3}_H = {\bf 1}_H +
{\bf 2}_H$ and ${\bf 3}_H' = {\bf 1}_H' + {\bf 2}_H'$ respectively.
The fragments ${\bf 2}_H$ and ${\bf 2}_H'$ are doublets of $SU(2)_W$
and have the quantum numbers of the electroweak doublets. The
non-zero vacuum expectation values (VEVs) are developed by the
singlet components, $ \langle {\bf 1}_H' \rangle \neq 0$ and  $
\langle {\bf 1}_H \rangle \neq 0$. As a result  of this breaking,
gauge bosons $V_{\mu}$ become massive. The following combination,
\begin{equation}
\label{true} {\bf 2}_{Gold} \, \equiv \,  {\langle {\bf 1}_H \rangle
{\bf 2}_H\, + \, \langle {\bf 1}_H' \rangle {\bf 2}_H' \over
\sqrt{\langle {\bf 1}_H \rangle^2 \, + \, \langle {\bf 1}_H'
\rangle^2 }} \,
\end{equation}
is eaten-up and becomes a longitudinal components of $V_{\mu}$. Whereas
the orthogonal state,
\begin{equation}
\label{higgs} {\bf 2}_{Higgs} \, \equiv\,  {\langle {\bf 1}_H'
\rangle {\bf 2}_H\, - \, \langle {\bf 1}_H \rangle {\bf 2}_H' \over
\sqrt{\langle {\bf 1}_H \rangle^2 \, + \, \langle {\bf 1}_H'
\rangle^2 }} \,
\end{equation}
is a pseudo-Nambu-Goldstone, which is massless at the tree-level and
gets the suppressed mass only at the loop level, and therefore
remains lighter than the symmetry breaking scale. This
pseudo-Goldstone plays the role of the Standard Model Higgs doublet.

\subsection{Goldstone Sister Higgs}

In the second approach \cite{sister}, the Higgs mass is protected
because it is related by symmetry to an {\it exact} Goldstone boson
that becomes a longitudinal component of $V_{\mu}$. We shall refer
to this scenario as ``Goldstone-Sister Higgs''. The  gauge symmetry
structure of the simplest model  is identical to the previous case.
There is an exact $U(3)_W$ gauge symmetry that incorporates the
Standard Model group as its maximal subgroup. Again, the spontaneous
breaking is triggered by two Higgs triplets, ${\bf 3}_H$ and ${\bf
3}_H'$. However, no approximate global symmetry is required.
Instead, the two Higgs triplets are related by an exact custodial
symmetry, such as, the permutation or an $SU(2)_{cust}$ symmetry
that transforms the triplets into each other ${\bf 3}_H
\rightleftharpoons {\bf 3}_H'$.

The lightness of the Higgs doublet is then guaranteed by the
following effect. Breaking of  $U(3)$ symmetry is triggered by the
VEV of the singlet component of the ${\bf 3}_H'$-triplet. During
this breaking, $V_{\mu}$ becomes massive and eats up the doublet
${\bf 2}_H'$. Thus ${\bf 2}_H'$ becomes a longitudinal polarization
of a massive gauge field
\begin{equation}\label{massV}
    V_{\mu} \, \rightarrow  \, V_{\mu} \,  + \,  {1 \over M_V}\,
    \partial_{\mu} \left({\bf 2}_H'\right),
\end{equation}
where $M_V$ is the mass of $V_{\mu}$. Since ${\bf 2}_H'$ is a true
eaten-up Goldstone, it cannot have any contribution to its mass from
the scalar potential, but only from the kinetic mixing with the
gauge field.  But, since the un-eaten doublet ${\bf 2}_H$ is related
to ${\bf 2}_H'$ by the custodial symmetry, the former also stays
massless  at the three-level. In this way the mass of the physical
Higgs doublet is protected by its sister doublet becoming a
Goldstone particle. We shall consider this scenario in more details
below.

\subsection{Higgs as Extra Dimensional Gauge Field}

Finally, the third class of theories in which appearance of the
doublet gauge fields is the must, is the one in which the Standard
Model Higgs doublet $H$ is an extra dimensional component of a
high-dimensional gauge field~\cite{olded,highd}. For understanding
the key idea of this approach,  it suffices to consider a simplest
case of a vector field in five dimensional Minkowski space, $V_{A}$,
where  $A = \mu, 5$ is the five dimensional Lorentz  index.   In the
approach of \cite{olded,highd}, the Higgs is identified with the
fifth component of the gauge field, which is a four-dimensional
Lorentz-scalar, $H \equiv V_5$.  Obviously,  since $H$ and $V_{\mu}$
are the components of the same high-dimensional gauge field, their
internal quantum numbers must be identical.

Thus, the existence of the weak-doublet  vector particles is
reinforced by the high-dimensional gauge symmetry. This symmetry is
spontaneously broken by compactification. In this way the mass of
the Higgs doublet is controlled by the compactification scale,  as
opposed to the high-dimensional cutoff of the theory. The realistic
model building in this class of theories is much more involved than
in the previous two cases.  For us the only important aspect is the
model-independent property of the existence of the massive vector
doublet $V_{\mu}$. This property  is guaranteed  by the symmetry and
is insensitive to the concrete model building.

Having specified the class of the theories of our interest, let us
turn to the interaction between $V_{\mu}$ and the Standard Model
fermions. In all three classes of theories, the coupling
(\ref{maincoupling}) even if not present  at the tree-level can (and
in general will) be generated by the loop corrections, as it is
permitted by all the symmetries of the low energy theory.  We shall
first demonstrate how this generation happens  by considering a toy
model reduced to its bare essentials, and later  illustrate it on a
detailed example of Goldstone-Sister Higgs \cite{sister}.

\section{Toy model}
In this section we shall discuss a generation of coupling
(\ref{maincoupling}) in a simple toy model. The latter
consists of two sectors: Hypothetical heavy particles and chiral
massless fermions of ordinary matter.
The corresponding Lagrangian
reads,
\begin{eqnarray}
\nonumber
  {\cal L} &=&  - D_{[\mu} V^\dagger_{\nu]} D^{[\mu} V^{\nu]}\, + \, M^2_V V^\dagger_\mu V^\mu
  +\partial_\mu\varphi^*\partial^\mu\varphi-M^2\varphi^*\varphi\\
\nonumber
  &+&\sum_{k=1,2}\bar{\psi}'_k \left(i D\!\bk-m\right)\psi'_k
  + g \bar{\psi}'_2\gamma^\mu\psi'_1 V_\mu
  +g V^\dagger_\mu\bar{\psi}'_1\gamma^\mu\psi'_2\\
\nonumber
  &+&\sum_{k=1,2}\bar{\psi}_k i D\!\bk \,\psi_k
  +\frac{h}{2}\bar{\psi}_2(1+\gamma^5)\psi'_2\varphi
  +\frac{h}{2}\bar{\psi}'_1(1+\gamma^5)\psi_1\varphi^*\\
  &+& \frac{h}{2}\bar{\psi}'_2(1-\gamma^5)\psi_2\varphi^*
  +\frac{h}{2}\bar{\psi}_1(1-\gamma^5)\psi'_1\varphi,
\end{eqnarray}
where the first line represents the bilinear Lagrangian of the
$SU(2)_W$-doublet vector fields, $V_\mu(V^\dagger_\mu)$, and of a
complex singlet scalar field, $\varphi$. The second line describes
the kinetic and mass terms of a singlet ($\psi'_1$) and a doublet
($\psi'_2$) heavy fermions and their interactions with the doublet
vector fields. The primed fermions are non-chiral, meaning that each
of them comes in both left and right chiralities. The last two lines
include the kinetic terms of the ordinary chiral massless fermions,
the left-handed doublets, $\psi_2$, and the right-handed singlets,
$\psi_1$, and their interactions with the heavy fields.

\begin{figure}[ht]
\epsfig{file=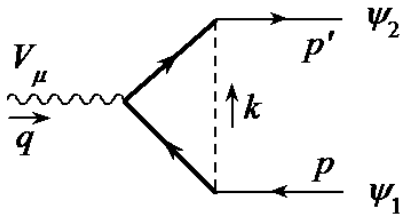,width=5.5cm} \caption{\label{fig:1}
Generation of new coupling.}
\end{figure}

The one-loop diagram in Fig.~\ref{fig:1}
leads to new anomalous coupling of the vector fields with the
ordinary fermions
\begin{eqnarray}
\nonumber
  \hspace{-0.2cm}\Delta{\cal L} &=& \frac{igh^2}{4}\bar{\psi}_2\int\frac{d^4k}{(2\pi)^4}
  (1+\gamma^5)\frac{p\bk'-k\bk+m}{(p'-k)^2-m^2}\gamma^\mu \\
\nonumber
   &&\times
   \frac{p\bk-k\bk+m}{(p-k)^2-m^2}(1+\gamma^5)\frac{1}{k^2-M^2}\psi_1
   V_\mu\\
   &=& \!\!\frac{gh^2}{32\pi^2 m}{\cal
   I}(q^2,m^2,M^2)\bar{\psi_2}\sigma^{\mu\nu}
   (1+\gamma^5)\psi_1\partial_{[\mu} V_{\nu]},
\end{eqnarray}
where
\begin{equation}\label{I}
    {\cal I}=\int_0^1 x^2dx \int_0^1\frac{ydy}{x+\frac{M^2}{m^2}(1-x)
    -\frac{q^2}{m^2}x^2y(1-y)}
\end{equation}
is a slow varied function at $q^2\ll m^2\sim M^2$.

\section{A model}

In this section we shall discuss an explicit example of the
realistic model.  As such we shall choose a model based on the idea
of Goldstone-Sister Higgs \cite{sister}. As discussed above, in this
theory the electroweak $SU(2)_W\times U(1)_Y$-symmetry is enhanced
to $U(3)_W$. Obviously the gauge sector contains an additional
electroweak-doublet vector field ($V_{\mu}$), which after
spontaneous breaking $U(3)_W\to SU(2)_W\times U(1)_Y$ becomes
massive. This breaking is realized by the VEV of a Higgs triplet
${\bf 3}_H'$, the weak-doublet part of which (${\bf 2}_H'$) becomes
a longitudinal component of $V_{\mu}$. In order to trigger the
second Standard Model stage of symmetry breaking $SU(2)_W\times
U(1)_Y \to U(1)_{EM}$, the theory contains a second Higgs triplet
${\bf 3}_H$. The key point is that the two triplets are related by
an additional custodial $SU(2)_{cust}$ symmetry. The motivation for
such symmetry is, that the physical Higgs doublet (${\bf 2}_H$) is a
partner of the Goldstone boson ${\bf 2}_H'$ that is eaten up via the
Higgs effect of $U(3)_W\to SU(2)_W\times U(1)_Y$ breaking. Due to
this symmetry relation, the SM Higgs doublet remains naturally
light.

It is the simplest to discuss this mechanism directly in
supersymmetric Grand Unified Theory (GUT) context, in which the
$U(3)_W \equiv SU(3)_W\times U(1)_W$ group together with the color
$SU(3)_C$ is embedded into the $SU(6)$ group as a maximal subgroup,
$G_{CW} \, \equiv\, SU(3)_C\times SU(3)_W\times U(1)_W\subset
SU(6)$. Notice that this embedding allows to treat color and weak
$SU(3)$-groups in a completely democratic way.

Notice, that for our present purposes having the full
$SU(6)$-symmetry group is completely unessential. The latter is
anyway broken down to its $G_{CW}$ subgroup at scales much above the
energies of our present interest. So we could have equally well
limit ourselves by $G_{CW}$ symmetry. However, the analysis is much
more convenient in terms of $SU(6)$ representations, rather than in
terms of its subgroups. It also allows us to understand the particle
content in term of representations of a more familiar $SU(5)$
subgroup.  Due to this advantages, we shall use
$SU(6)$-classification for the particles. If needed, the reader can
easily perform decompositions into the $G_{CW}$-reduced
representations instead.

The full symmetry of the model is thus $SU(6)\times SU(2)_{\rm
cust}$, where $SU(2)_{\rm cust}$ is an additional {\em custodial}
symmetry that relates the SM Higgs doublet to an {\em eaten-up
Goldstone boson}.

The chiral superfield content is:
\begin{enumerate}
  \item Higgs sector:\\ 35-plet $\Sigma^j_i~(i,j=1,\dots,6)$ and\\
  $({\bf 6}.{\bf 2})\equiv H_{Aj}$, $({\bf{\bar{6}}}.{\bf 2})\equiv \bar{H}^{Aj}$, where
 $i,j$ are $SU(6)$ and $A=1,2$ are $SU(2)_{\rm cust}$ indexes respectively.
  \item The SM fermions are embedded in the following anomaly-free
  set (per generation):\\
  ${\bf 15}_{[ij]}$, $({\bf{\bar{6}}}.{\bf 2})\equiv
  {\bf{\bar{6}}}^{Aj}$ and a singlet ${\bf 1}$.
\end{enumerate}

We shall denote the superfields by the same symbols as their
components. In each case it will be clear from the context which
component we are referring to.

The symmetry breaking is achieved by the Higgs part of the
superpotential, which has the following form,
\begin{equation}\label{WHiggs}
    W_{\rm Higgs}=\frac{\lambda}{3}{\rm Tr}\Sigma^3
    +\lambda'\bar{H}^A\Sigma H_A+M'\bar{H}^A H_A.
\end{equation}
The vacuum of the theory is:
\begin{equation}\label{vacuum}
    \Sigma=\!\left(
             \begin{array}{ccclll}
               1 &   &   &   &   &   \\
                 & 1 &   &   &   &   \\
                 &   & 1 &   &   &   \\
                 &   &   &-1 &   &   \\
                 &   &   &   &-1 &   \\
                 &   &   &   &   &-1 \\
             \end{array}
           \right)\!\frac{M'}{\lambda'},~
    \bar{H}^1=H_2=\left(
                    \begin{array}{c}
                      0 \\
                      0 \\
                      0 \\
                      0 \\
                      0 \\
                      \mu \\
                    \end{array}
                  \right),
\end{equation}
which leads to the symmetry breaking $SU(6)\to SU(3)_C\times SU(2)_W\times
U(1)_Y$. The value of $\mu$ is undetermined in SUSY limit.

We are interested in the situation when the scale $\mu$ is not very
high, around TeV or so, whereas the scale $M'/\lambda'$ is much
larger. In this way, below the latter scale the low energy symmetry
group is $G_{CW}$. The further breaking of this symmetry to the
Standard Model group, is triggered at much lower scale $\mu$ by the
VEV of the $6$-plets.

As a result of this breaking, the Higgs doublets in $\bar{H}^1$ and
$H_2$ are eaten up by the $SU(2)_W$-doublet gauge fields $V_{\mu}
(V_{\mu}^\dagger)$ that reside in the adjoint representation of
$SU(3)_W$. The doublets in $\bar{H}^2$ and $H_1$ are physical and at
the tree level (in SUSY limit) have very small masses $M\sim
\lambda\mu^2/M'$. An additional contribution comes from loop
corrections after supersymmetry breaking (see \cite{sister} for
details).

Notice, that as a result of the symmetry structure of the theory,
the color triplet partners of the Higgs doublets are automatically
super-heavy, with the masses $2M'$, and decouple from the low energy
spectrum. This solves the doublet-triplet splitting problem in SUSY
GUTs.

The fermion masses are generated from the following interactions.
The up-type quark masses come from
\begin{equation}\label{up}
    \varepsilon^{AB}H_{Ai}H_{Bj}{\bf 15}_{[km]}{\bf 15}_{[nl]}~\varepsilon^{ijkmnl},
\end{equation}
and the masses of down-type quarks, charged leptons and heavy states
come from
\begin{equation}\label{down}
    \varepsilon_{AB}\bar{H}^{Ai}{\bf 15}_{[ij]}{\bf{\bar{6}}}^{Bj}
    +H_{Aj}{\bf{\bar{6}}}^{Aj} {\bf{1}}.
\end{equation}

\begin{figure}[ht]
\epsfig{file=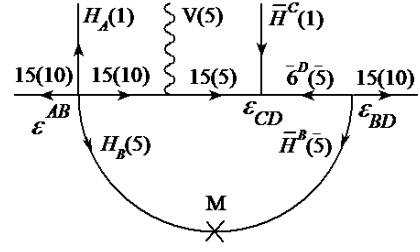,width=5.5cm} \caption{\label{fig:s1} Generation
of new coupling (\ref{Kup}). The numbers in the brackets refer to
the relevant representations with respect to the $SU(5)$ subgroup.
The appropriate supersymmetry-breaking insertions in various
vertexes are assumed.}
\end{figure}

\begin{figure}[ht]
\epsfig{file=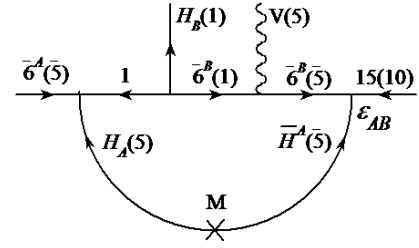,width=5.5cm} \caption{\label{fig:s2} Generation
of new coupling (\ref{Kdown}). The numbers in the brackets refer to
the relevant representations with respect to the $SU(5)$ subgroup.
The appropriate supersymmetry-breaking insertions in various
vertexes are assumed.}
\end{figure}

From the above couplings the following interactions between the
doublet vector and light fermions are generated. The superdiagram
containing external light up-type quarks is given in
Fig.~\ref{fig:s1}.
The corresponding effective operator of interest is,
\begin{equation}\label{Kup}
    \bar{H}^*_{aB}  (F_{\mu\nu})^a_j \, H_{Ai} \,
    {\bf 15}_{[km]}\sigma^{\mu\nu} {\bf 15}_{[ln]} \,
    \varepsilon^{BA}\varepsilon^{jikmln},
\end{equation}
where $(F_{\mu\nu})^a_j$ is the $SU(6)$ gauge field strength, the
fermionic components are taken from $15$-plets and bosonic
components from the rest. After substituting the VEVs of $6$-plet
Higgses, this operator reduces to the magnetic coupling between
$V_{\mu}$ and the up-type quarks given by the last term in
(\ref{maincoupling}). Coupling to down-type quarks and leptons is
generated through the diagram of Fig.~\ref{fig:s2},
which leads to the following effective operator,
\begin{equation}\label{Kdown}
    H^{*Ba}(F_{\mu\nu})_a^i \,
    {\bf 15}_{[ij]}  \sigma^{\mu\nu} {\bf \bar{6}}^{Aj}\varepsilon_{BA}.
\end{equation}
After substitution of the VEVs, the above interaction reduces to the
magnetic coupling of $V_{\mu}$ with $d$-quark and charged leptons,
given by the first two terms in (\ref{maincoupling}).

Notice, that due to supersymmetry violating insertions in the
vertexes, the flavor structure of the operators (\ref{Kup}) and
(\ref{Kdown}) is not necessarily aligned with the flavor structure
of quark and lepton mass matrixes. This means, that the exchange by
$V_{\mu}$ could potentially contribute into new flavor and CP
violating interactions.

Notice also, that the same operator (\ref{Kdown}) could potentially
give a contribution to $(g-2)$ of leptons. Indeed, if we instead of
inserting the $SU(2)_W$-doublet VEV of $H_2$, insert the VEV of a
physical Higgs doublet living in $H_1$, the operator (\ref{Kdown})
will reduce to the magnetic moment coupling of photon. However, this
magnetic coupling will mix the light left-handed lepton residing in
$15$-plet with the heavy right-handed lepton from ${\bf \bar{6}}^2$,
as opposed to the Standard Model right-handed lepton residing in
${\bf \bar{6}}^{1}$, thus, giving no contribution to $(g-2)$.
However, a non-zero contribution could arise in case of a small
mixing between the doublets from $H^1$ and $H^2$. Such mixing could
arise from the calculable radiative corrections, and thus, could
relate the phenomenological signatures of $V_{\mu}$ with the value
of $(g-2)$.

This study will not be attempted in the present work. Instead, we
shall focus on characteristic signatures of resonance production and
decay of $V_{\mu}$-bosons.

\section{Consequences for colliders}
In paper \cite{two} it has been shown that tensor
current$\times$current interaction leads to a new angular
distribution in comparison with well-known vector interactions. It
was realized later~\cite{misha} that this property ensures
distinctive signature for their detection at the hadron colliders.

The hadron colliders, due to their biggest center-of-momentum (CM)
energy $\sqrt{s}\sim$ several TeVs and their relatively compact
sizes, still remain the main tools for discoveries of very heavy
particles. The  presence of partons with a broad range of different
momenta allows to flush the entire energetically accessible region,
roughly, up to $\sqrt{s}/6$. The production mechanism for new heavy
bosons at a hadron collider is the $q\bar{q}$ resonance fusion.

In this paper we will consider the resonance production and decay of
the above-introduced heavy spin-1 gauge bosons into the light lepton
pairs, electrons or muons. For such high energies it is convenient
to use the helicity formalism, since the helicity is a good quantum
number for massless particles.

This, in a way, fixes the dominant production and decay mechanisms.
For example, the decay angular distribution in the CM frame of a
particle with spin-$s$ and helicity $\lambda$ with
$-s\leq\lambda\leq s$ decaying into two massless particles with
helicities $\lambda_1$ and $\lambda_2$ can be written
as~\cite{Haber}
\begin{equation}\label{ds/dcos}
    \frac{{\rm d}\Gamma_s}{{\rm d}\cos\theta\;{\rm d}\phi}=
    \frac{1}{64\pi^2 M}\vert{\cal
    M}^{s\lambda}_{\lambda_1\lambda_2}(\theta,\phi)\vert^2,
\end{equation}
where the helicity amplitude
\begin{equation}\label{amplitude}
    {\cal M}^{s\lambda}_{\lambda_1\lambda_2}(\theta,\phi)=
    \sqrt{\frac{2s+1}{4\pi}}e^{i(\lambda-\delta)\phi}
    d^s_{\lambda\delta}(\theta){\cal M}^s_{\lambda_1\lambda_2}
    \vspace{0.2cm}
\end{equation}
is expressed through the difference
$\delta\equiv\lambda_1-\lambda_2$ and the reduced decay amplitude
${\cal M}^s_{\lambda_1\lambda_2}$, which is a function of $s$ and
the outgoing helicities, but is independent of the polar  ($\theta$)
and the azimuthal  ($\phi$) angles.

Up until now only resonance production and decay of spin-1 bosons
with maximal helicities $\lambda=\pm 1$ have been considered. They
are associated with additional $U(1)'$ gauge symmetries and are
usually called $Z'$ particles. The Lorentz structure of its couplings to
each fermion flavor is characterized by two generally independent
constants $g^f_{LL}$ and $g^f_{RR}$
\begin{equation}\label{LL}
    {\cal L}_{Z'}=\sum_f \left(g^f_{LL}\;\overline{\psi^f_L}\gamma^\mu\psi^f_L
    +g^f_{RR}\;\overline{\psi^f_R}\gamma^\mu\psi^f_R\right) Z'_\mu \, .
\end{equation}
Experimental determination of these coupling constants or
disentangling among the different models is a rather hard task and
cannot be fulfilled, for example, with the first LHC data. In the
best case only the specific symmetric angular distribution over the
polar angle $\theta$,
\begin{equation}\label{GLL}
    \frac{{\rm d} \Gamma_1(q\bar{q}\to Z'\to\ell\bar{\ell})}
    {{\rm d} \cos\theta} \propto
    \vert d^1_{11}\vert^2+\vert d^1_{-11}\vert^2 \sim
    1+\cos^2\theta \, ,
\end{equation}
allows to distinguish its production from distributions of spin-0,
\begin{equation}\label{G0}
    \frac{{\rm d} \Gamma_0(q\bar{q}\to h\to\ell\bar{\ell})}
    {{\rm d} \cos\theta} \propto \vert d^0_{00}\vert^2\sim 1 \,
\end{equation}
and spin-2 \,
\begin{eqnarray}\label{G2qq}
    \frac{{\rm d} \Gamma_2(q\bar{q}\to G^*\to\ell\bar{\ell})}
    {{\rm d} \cos\theta} &\propto& \vert d^2_{11}\vert^2
    +\vert d^2_{-11}\vert^2
    \nonumber\\
    &\sim& 1-3\cos^2\theta+4\cos^4\theta\\
    \label{G2gg}
    \frac{{\rm d} \Gamma_2(gg\to G^*\to\ell\bar{\ell})}
    {{\rm d} \cos\theta} &\propto& \vert d^2_{21}\vert^2
    +\vert d^2_{-21}\vert^2
    \nonumber\\
    &\sim& 1-\cos^4\theta \,
\end{eqnarray}
resonances.

Another possibility is the resonance production and decay of
longitudinal spin-1 bosons with $\lambda=0$, but  this possibility
is not widely discussed.

While the $Z'$ bosons with helicity $\lambda=\pm 1$ are produced in
left(right)-handed quark and right(left)-handed antiquark fusion,
the longitudinal $Z^*$ bosons can be produced through the new chiral
couplings (\ref{maincoupling}),
\begin{equation}\label{LR}
    {\cal L}_{Z^*}\!=\!\!\!\sum_{f=d,e}\!\!\left(\frac{g^f_{LR}}{M}
    \overline{\psi^f_L}\sigma^{\mu\nu}\psi^f_R\partial_{[\mu}
    \bar{Z}^*_{\nu]}
    +\frac{g^f_{RL}}{M}\overline{\psi^f_R}\sigma^{\mu\nu}
    \psi^f_L\partial_{[\mu} Z^*_{\nu]}\right)\!,
\end{equation}
with the complex constants $g^f_{LR}=(g^f_{RL})^*$ in left-handed or
right-handed quark-antiquark fusion~\cite{proposal}.

The new couplings lead to a different angular distribution
\begin{equation}\label{GLR}
    \frac{{\rm d} \Gamma_1(q\bar{q}\to Z^*\to\ell\bar{\ell})}
    {{\rm d} \cos\theta} \propto
    \vert d^1_{00}\vert^2\sim\cos^2\theta,
\end{equation}
than the previously considered ones. At first sight, the small
difference between the distributions (\ref{GLL}) and (\ref{GLR})
seems unimportant. However, the absence of the constant term in the
latter case results at least in two potential  experimental
signatures.

\begin{figure}[h]
\centering \epsfig{file=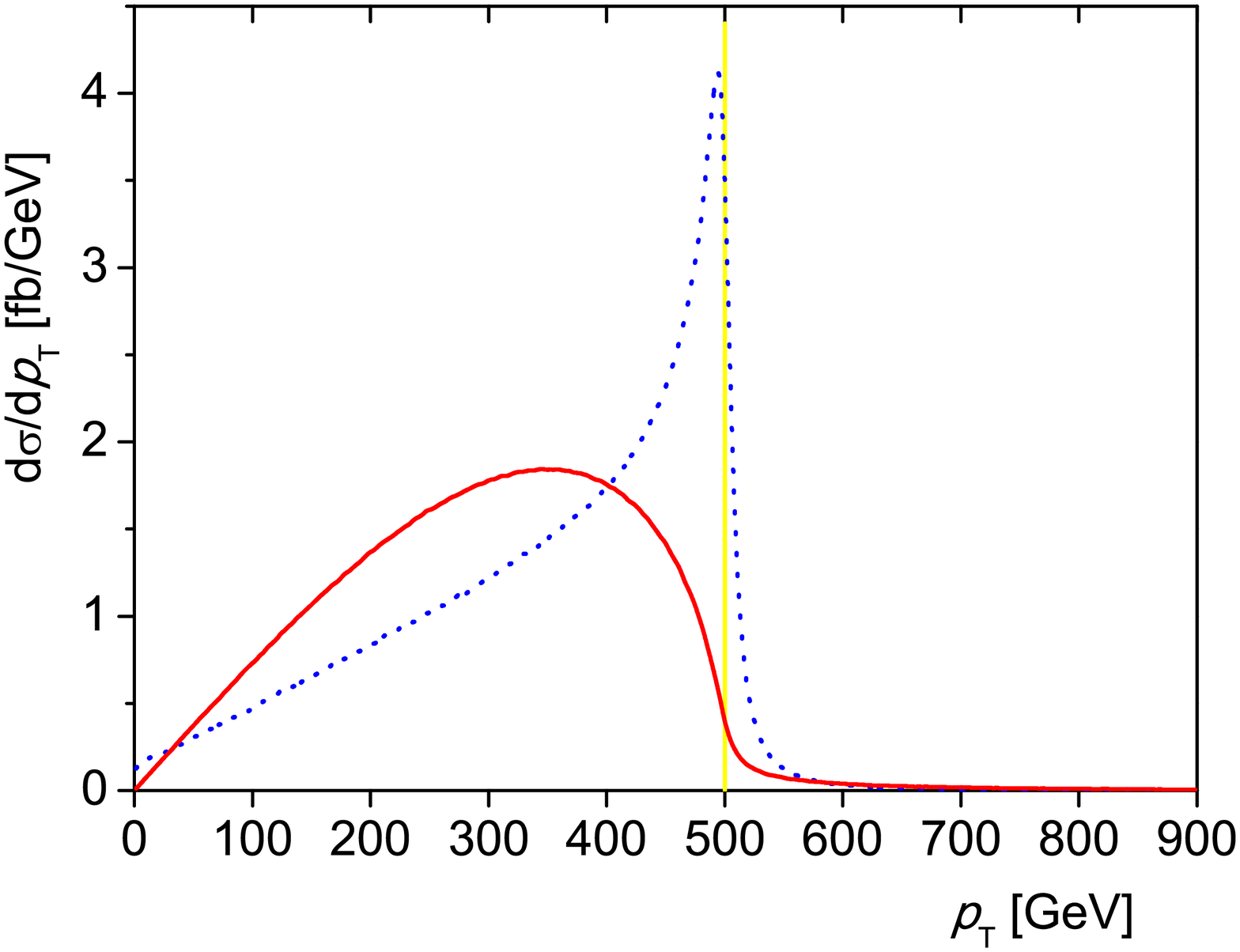,width=0.48\textwidth}
\caption{\label{fig:pt} The lepton $p_T$ distributions from the
$Z^*$ (solid) and $Z'$ (dotted) bosons decays.}
\end{figure}

First of all, the known angular distributions for scalar (\ref{G0}),
vector (\ref{GLL}) and spin-2 (\ref{G2qq},\ref{G2gg}) bosons include
a nonzero constant term, which leads to the kinematic singularity in
$p_T$ distribution of the final lepton
\begin{equation}\label{1/cos}
    \frac{1}{\cos\theta}\propto\frac{1}{\sqrt{(M/2)^2-p^2_T}}
\end{equation}
in the narrow width approximation $\Gamma <\!\!\!< M$
\begin{equation}\label{narrow}
    \frac{1}{(s-M^2)^2+M^2\Gamma^2}\approx\frac{\pi}{M\Gamma}\delta(s-M^2).
\end{equation}
This singularity is transformed into a well known Jacobian peak due
to a finite width of the resonance. In contrast to this, the pole in
the decay distribution of the $Z^*$ bosons is canceled out and the
lepton $p_T$ distribution even reaches zero at the kinematical
endpoint $p_T=M/2$. Therefore, the $Z^*$ boson decay distribution
has a broad smooth hump with the maximum below the kinematical
endpoint, instead of a sharp Jacobian peak (Fig.~\ref{fig:pt}).

\begin{figure}[h]
\centering \epsfig{file=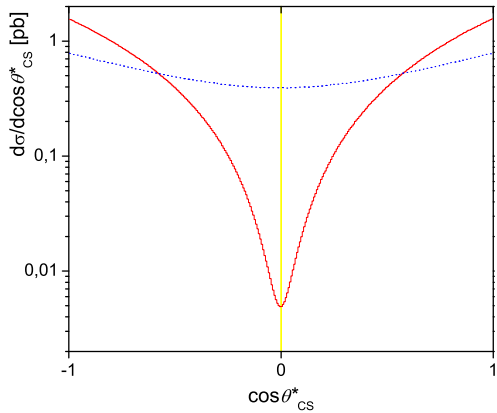,width=0.43\textwidth}
\caption{\label{fig:cos} The lepton angular distributions from the
$Z^*$ (solid) and $Z'$ (dotted) bosons decays.}
\end{figure}

Another striking feature of the distribution (\ref{GLR}) is the
forbidden decay direction perpendicular to the boost of the excited
boson in the rest frame of the latter (the Collins--Soper
frame~\cite{CS}). It leads to a peculiar ``swallowtail'' shape of
the angular distribution with a profound dip at $\cos\theta^*_{\rm
CS}=0$ in the Collins--Soper frame
(Fig.~\ref{fig:cos})~\cite{proposal}.

In conclusion, we would like to emphasize the difference between
group properties of the gauge spin-1 $Z'$ and $Z^*$ bosons. While
the $Z'$ bosons are described by real representations transforming
as triplets or singlets  of $SU(2)_W\times U(1)_Y$ group, the $Z^*$
bosons, like the Higgs fields, are assigned to the complex
representation and transform as doublets.  This results in another
important experimental consequence.

Together with the neutral $Z^{*}$ bosons the doublets always contain
also the charged bosons, $W^{*\pm}$, which decay into a charged
lepton and an undetected neutrino. Therefore, the angular
distribution in Fig.~\ref{fig:cos} is experimentally unaccessible
for them in the lepton channel.  Only the $p_T$-distribution of the
charged lepton can be measured. However, the distribution of the
$W^*$ bosons differs drastically from the distribution of the $W'$
bosons (Fig.~\ref{fig:pt}). Hence, even relatively small decay width
of the $W^*$ bosons leads to a wide hump without the Jacobian peak,
that obscures their identification as resonances at the hadron
colliders.

The only way to access the angular distribution for $W^*$ bosons
like in Fig.~\ref{fig:cos} is kinematical reconstruction of their
decays into heavy quarks, $t\bar{b}$, which, in spite of the strong
QCD background,  can be identified via $b$-tagging. The presence of
the heaviest $t$ quark in the final state will lead to an additional
contribution,
\begin{equation}\label{d110}
    \vert d^1_{01}\vert^2\sim 1-\cos^2\theta \, ,
\end{equation}
to the angular distribution (\ref{GLR}) proportional to the ratio
$m^2_t/M^2$ due to helicity flip of the $t$ quark.

\section*{Acknowledgements}
We are grateful to C. Wetterich for discussions. The work of MC was
partially supported by Grant-in-Aid for Scientific Research
104/15.05.2009 from the Sofia University. The research of GD is
supported in part by European Commission under the ERC advanced
grant 226371, by David and Lucile Packard Foundation Fellowship for
Science and Engineering and by the NSF grant PHY-0758032.


\end{document}